**Quantum textures of the many-body wavefunctions in magic-angle graphene**

Kevin P. Nuckolls[1,*], Ryan L. Lee[1,*], Myungchul Oh[1,*], Dillon Wong[1,*], Tomohiro Soejima[2,*], Jung Pyo Hong[1], Dumitru Călugăru[1], Jonah Herzog-Arbeitman[1], B. Andrei Bernevig[1,3,4], Kenji Watanabe[5], Takashi Taniguchi[6], Nicolas Regnault[7], Michael P. Zaletel[2,8], Ali Yazdani[1,†]

*[1] Joesph Henry Laboratories and Department of Physics, Princeton University, Princeton, NJ 08544, USA*

*[2] Department of Physics, University of California, Berkeley, Berkeley, CA 94720, USA.*

*[3] Donostia International Physics Center, P. Manuel de Lardizabal 4, 20018 Donostia-San Sebastian, Spain*

*[4] IKERBASQUE, Basque Foundation for Science, Bilbao, Spain*

*[5] Research Center for Functional Materials, National Institute for Materials Science, 1-1 Namiki, Tsukuba 305-0044, Japan*

*[6] International Center for Materials Nanoarchitectonics, National Institute for Materials Science, 1-1 Namiki, Tsukuba 305-0044, Japan*

*[7] Laboratoire de Physique de l'Ecole normale supérieure, ENS, Université PSL, CNRS, Sorbonne Université, Université Paris-Diderot, Sorbonne Paris Cité, 75005 Paris, France*

*[8] Materials Sciences Division, Lawrence Berkeley National Laboratory, Berkeley, CA 94720, USA.*

* These authors contributed equally to this work.
† Corresponding author email: yazdani@princeton.edu

**Interactions among electrons create novel many-body quantum phases of matter with wavefunctions that often reflect electronic correlation effects, broken symmetries, and novel collective excitations. A wide range of quantum phases has been discovered in magic-angle twisted bilayer graphene (MATBG), including correlated insulating[1], unconventional superconducting[2–5], and magnetic topological phases[6–9]. The lack of microscopic information[10,11], including precise knowledge of possible broken symmetries, has thus far hampered our understanding of MATBG's correlated phases[12–17]. Here we use high-resolution scanning tunneling microscopy to directly probe the wavefunctions of the correlated phases in MATBG. The squares of the wavefunctions of gapped phases, including those of the correlated insulators, pseudogap, and superconducting phases, show distinct patterns of broken symmetry with a √3 x √3**

**super-periodicity on the graphene atomic lattice that has a complex spatial dependence on the moiré superlattice scale. We introduce a symmetry-based analysis to describe our measurements of the wavefunctions of MATBG's correlated phases with a set of complex-valued local order parameters. These order parameters and their phases show intricate textures, including vortices associated with the order parameters' phase windings, which distinguish the various correlated phases. For the correlated insulators in MATBG, at fillings of v = ±2 electrons per moiré unit cell relative to charge neutrality, we compare the observed quantum textures to those expected for proposed theoretical ground states. In typical MATBG devices, the textures of correlated insulators' wavefunctions closely match those of the theoretically proposed incommensurate Kekulé spiral (IKS) order[15], while in ultra-low-strain samples our data has local symmetries like those of a time-reversal symmetric intervalley coherent (T-IVC) phase[12]. We also study the wavefunction of MATBG's superconducting state, revealing strong signatures of intervalley coherence that can only be distinguished from those of the insulator with our phase-sensitive measurements.**

Magic-angle twisted bilayer graphene (MATBG), created by stacking and twisting two layers of graphene to 1.1°, is an archetype of a new class of quantum materials in which moiré patterns give rise to electronic bands with flat energy dispersion. At partial filling of its flat bands, when electronic correlations are strong, MATBG exhibits insulating[1], unconventional superconducting[2–5], and magnetic topological phases[6–9]. Of particular interest is the nature of the correlated insulators at fillings of v = ±2 (two electrons or holes per moiré unit cell), since an unconventional superconducting phase emerges from doping these insulators. Theoretical works have proposed several candidate ground states for these insulators[12,13,15,14], but these states have similar calculated energies, and experiments to date have not provided sufficient information to distinguish among them. The recent success of scanning tunneling microscopy / spectroscopy (STM / STS) in studying the broken symmetry states of graphene's zeroth Landau level[18,19] has motivated theoretical studies to propose distinguishable signatures among various ground states that reflect their broken symmetries in the local density of states (LDOS)[10,11].

Here, we report STM measurements as a function of filling of the flat bands of MATBG, showing that the opening of correlation-induced gaps, either for the correlated insulating phases at v = ±2 or for superconducting or pseudogap phases at nearby densities, coincide with a (√3 x √3) tripling of the unit cell detected in the LDOS on the graphene atomic lattice scale, which we call the "R3 pattern". The R3 pattern indicates the presence of intervalley coherent (IVC) order[10,11], but these patterns vary spatially on the moiré-scale in different ways among the correlated phases. To characterize these quantum textures, we use a symmetry-based analysis of high-resolution STM data in MATBG obtained over large fields of view (up to 60,000 atoms). Our measurements reveal moiré-scale translation- and rotational-symmetry-breaking that depend on the degree of strain and doping in our devices[20]. This finding, together with observations of complex electronic wavefunction textures such as vortex-like features, allows us to show that the predicted incommensurate Kekulé spiral (IKS) states[15] are likely the ground states at ν = ±2 in typical devices. We also distinguish quantum wavefunctions of the correlated insulators in typical versus ultra-low-strain devices, and show how they evolve when the sample is in the superconducting phase at non-integer filling. Our work represents an unprecedented measurement of correlated many-body wavefunctions in a pristine material system, allowing us to identify remarkably complex quantum states that describe the correlated insulating phases of MATBG. Our measurements of the superconducting phase at nearby densities also place important constraints on models of pairing in MATBG[21–24].

We examined the spatial structure of the electronic states with high spatial and energy resolution in gated MATBG devices using a home-built dilution fridge STM[25] (at 4 K and 200 mK) as a function of filling of the flat bands (see Methods for details of measurements and device fabrication). Our experiments were made possible by fabricating ultra-clean samples in which large (> 100 x 100 $nm^2$) regions were easily found without any local scatterers or residual polymer from the fabrication process, combined with a well-established tip calibration procedure. Examining a representative low-bias STM topograph of the correlated insulating phase at ν = +2 (Fig. 1b), we find the LDOS of this phase shows real-space atomic-scale features indicative of symmetry-breaking on the graphene lattice scale, which triples the graphene unit cell. Early STM studies of MATBG[26–28] reported that the LDOS near AA sites develops asymmetric features in the correlated insulating phase; however, we do not find any such features in our high-resolution atomically resolved measurements on ultra-clean devices using well-calibrated tips (see SI).

To correlate the observations of this atomic-scale symmetry breaking with the formation of correlated insulating phases near ν = ±2, we examine measurements of such patterns for different fillings of the flat bands near ν = -2. As in previous studies[5], STS data in Fig. 1c shows a correlation-induced gap indicative of an insulating phase near ν = -2. As shown in Fig. 1d, we find that for fillings away from ν = -2, low-bias STM images across the moiré superlattice show the atomic lattice of the top layer of MATBG. However, for fillings near the correlated insulator at ν = -2, low-bias images of both occupied and unoccupied states show broken symmetry features on the atomic-scale, with the strongest intensity at $\nu = -2$. To quantify our observations, we correlate the information obtained from the Fourier transform (FFT) of large field of view images (40 x 40 $nm^2$) at different fillings shown in Fig. 1d with the spectroscopic data of Fig 1c. As the carrier density approaches the correlated insulator filling, we observe peaks at $Q_{IVC}$ (blue diamond markers) at a distance of $4\pi/\sqrt{3}a_0$ ($a_0$ is the graphene lattice constant) in addition to the Bragg peaks of the graphene lattice $Q_{Bragg}$ (black hexagon markers). $Q_{IVC}$ wavevectors connect the graphene valleys at **K** and **K'**, indicating that the wavefunction at $\nu = \pm 2$ is a coherent superposition of states in the two valleys. The tripling of the unit cell, i.e. the R3 pattern, is most intense at $\nu = -2$, where we also observe their associated second-order peaks $2Q_{IVC}$ (blue octagon markers). $\Sigma\,|FFT(Q_{IVC})|\ /\ \Sigma\,|FFT(Q_{Bragg})|$ provides a measure of the strength of the broken symmetry R3 patterns, which we plot together with the zero-bias tunneling resistance next to the STS measurements as a function of density. We find these quantities to be perfectly coincident (see SI for complete data set). We have performed several experimental checks to confirm that the R3 patterns are intrinsic to the strongly correlated phases of MATBG, such as finding that they are absent in non-magic angle devices, absent when the chemical potential lies outside the flat bands, and are unrelated to any local scatterers[29] (see SI for extensive details).

We find the R3 pattern to be a salient feature of the wavefunction of several correlated phases of MATBG; however, the nature of these wavefunctions is far more complex than the simple presence of such symmetry-broken patterns. This complexity is revealed when we examine LDOS in ultra-clean devices over many moiré unit cells, an example of which for $\nu = -2$ is shown in Fig. 2a. While the real-space maps and their FFTs both show R3 ($Q_{IVC}$) symmetry-breaking features at all locations, the spatial patterns of such symmetry-breaking can be dramatically different from one location to another within the moiré unit cell (right panels; Fig. 2a). This change is also reflected in both the amplitude and the phase of the $Q_{IVC}$ peaks of these local regions (Fig. 2a,b). Moreover, in different devices, we find that the R3 pattern can be different between equivalent locations in the moiré superlattice in different samples. These variations, and those we study in detail below as a function of carrier density in the same device, motivate us to perform a systematic symmetry-based analysis of local STM data that more precisely characterizes the features of the wavefunctions of these correlated phases.

Our symmetry analysis decomposes the STM data into a minimal set of local order parameters that capture the transformation properties of the R3 pattern under the symmetries of the graphene lattice. STM images of the correlated states in MATBG are described by six independent complex-valued numbers, derived from the first-order $Q_{Bragg}$ peaks and the $Q_{IVC}$ peaks of the R3 pattern (STM images are real-valued, so the Fourier transform at **k** and -**k** are not independent). These complex numbers can be decomposed using the irreducible representations of the $C_3$ point group into six corresponding complex-valued order parameters (see SI), three deriving from $Q_{Bragg}$, which capture the degree of sublattice polarization ($C_2$-breaking) and lattice-scale nematicity ($C_3$-breaking), and three deriving from $Q_{IVC}$, which describe the various types of R3 patterns. In the vicinity of AA-stacked regions of the moiré superlattice, the latter three capture the intensity of the R3 pattern on graphene's A sublattice, B sublattice, or on the bonds of the graphene lattice, which we refer to respectively as the "IVC Site A", "IVC Site B", and "IVC Bond" order parameters. Emblematic examples of LDOS patterns with specific values of these order parameters are shown in Figs. 2c-d, which also illustrate how the R3 patterns change as we vary the phases of the order parameters.

We can get some physical intuition for these order parameters and their connection to symmetry-breaking and valley isospin-ordering by considering their relation to the so-called chiral model of MATBG[16], which has eight degenerate flat bands arising from their valley $\tau^z$ = (**K**, **K'**), sublattice $\sigma^z$ = (A, B), and spin $s^z$ = (↑, ↓) indices. Each of the bands carries either Chern number $\mathbb{C}$ = +1 or –1 according to the product of its valley and sublattice indices: $\mathbb{C} = \tau^z \cdot \sigma^z$. With this identification, it is natural to instead label states by ($\mathbb{C}$, $\tau^z$, $s^z$) so that wavefunctions within each independent C = ±1 sector (each like the zeroth Landau level in monolayer graphene[18,19]) can be represented by an isospin vector ($\tau^x$, $\tau^y$, $\tau^z$) on a Bloch sphere describing states from valley polarized (vector up or down) to intervalley coherent (vector along the equator). The patterns in Fig. 2c, with LDOS spectral weight on carbon-carbon bonds generated by the IVC Bond order parameter, can be understood as adding the densities of two states described by $\mathbb{C}$ = ±1 Bloch sphere vectors that lie along the equatorial plane. This is because the rule $\mathbb{C} = \tau^z \cdot \sigma^z$ implies that intra-Chern, intervalley coherence occurs between opposite sublattices (e.g. R3 pattern centered on bonds).

However, the three patterns with nonzero IVC Site A/B order parameter shown in Fig. 2d represent novel states that cannot be generated by simple addition of densities from the two Chern sectors. They represent inter-Chern-sector coherent states described by a third Bloch sphere prescribing how electronic states between these independent Chern sectors coherently interfere. This inter-Chern, intervalley coherence hybridizes states within the same sublattice, leading to an R3 pattern centered on either the A or B sites. The appearance of such patterns in our experimental data reveals this novel feature of physics in MATBG, which can give rise to more complex types of quantum Hall ferromagnet-like phases forming across Chern sectors. In total, the spatial patterns in the data signal the complexity of intervalley and inter-Chern coherent phases that can form in MATBG, which require a precise method of experimental characterization.

The experimentally observed R3 patterns can be described by combining the patterns of the IVC-Bond, IVC-A, and IVC-B order parameters (Fig. 2b). To perform a precise analysis, we use the $Q_{IVC}$ peak amplitudes and phases in the FFTs of subregions (1 x 1 $nm^2$) of atomically resolved large field of view STM images (50 x 50 $nm^2$) to construct maps of the IVC order parameters, and study them as a function of doping between different correlated states. Decomposition of the experimental data into the IVC order parameters requires making the complex phase of each local FFT well-defined across large fields of view, a task that requires locating the centers of all graphene hexagons in STM images. Details of the procedure for hexagon finding and the order parameter extraction from STM data are described in the SI section.

We apply our order parameter analysis to STM images of the commonly observed correlated insulating phase at $v = -2$ in a typical MATBG device at T = 200 mK. Typical, strained MATBG devices ($\varepsilon \sim 0.1 - 0.4\%$) display gapless, semi-metallic behavior at charge neutrality ($v = 0$, Fig. 3a), as reported in previous studies[5,30]. Across a large field of view (~50 x 50 $nm^2$), Fig. 3b shows a grayscale plot of the real-valued total IVC strength and color plots of the complex-valued IVC Bond, IVC Site A, and IVC Site B order parameters, where the color represents the phase of these order parameters while the brightness represents their magnitude at each point. We find the IVC Total map, which measures the strength of the R3 pattern, to be periodic on the moiré-scale. In strong contrast to this moiré-periodic behavior, we find the IVC-A, IVC-B, and IVC-Bond order parameters break moiré translation and rotational symmetries, displaying stripe-like features that show broad regions of roughly constant phase separated by narrow regions in which the phase changes suddenly. Moreover, we observe vortices and antivortices (phase winding $\Delta\theta = \pm 2\pi$ over a closed loop) in the IVC order parameter maps upon subtracting a linear phase background, used simply to aid the visual identification of vortices without affecting their locations (Fig. 3b, see SI for more details). The phase information (Fig. 3b and zoom-in 3e) contains certain vortices and antivortices that appear with the same chirality across the three different IVC order parameters, with cores centered near deep suppressions of the total IVC strength (into and out-of-plane markers). The maps also contain other vortices that appear in featureless locations in the IVC Total plots and are uncorrelated between different IVC order parameters (black dot markers). The observed moiré-scale symmetry-breaking and vortex features cannot be detected through simple visual examination of STM images or FFT amplitudes, since the key information for uncovering these phenomena is contained in the relative FFT phases between different locations.

We uncover an important aspect of the correlated insulating phases in MATBG by contrasting our findings discussed above in typical strained (ε ~ 0.1% - 0.4%) MATBG samples with those of ultra-low-strain (ε < 0.1%) MATBG samples shown in Figs. 3c-d. In this sample (ε = 0.03%) and other rarely found ultra-low-strain device regions, we find that the LDOS at charge neutrality is gapped (Fig. 3c) and convolved with Coulomb charging effects (see SI). From this observation, even in the absence of hBN alignment[6,7,5,9], we infer that strong interactions can gap the Dirac points of low-strain MATBG. The correlated insulating phase at v = -2 in this sample (Fig. 3d) also shows R3 patterns whose strength is moiré-periodic, with maxima at the AA sites. However, in strong contrast to the typically strained MATBG devices, Fig. 3d shows that the magnitudes and phases of the IVC-Bond, IVC-A, and IVC-B order parameters are all nearly moiré-periodic in this ultra-low-strain sample. Furthermore, subtracting a linear phase background from these maps also shows behavior distinct from those in typical samples, with the absence of any vortex-antivortex features in the nearly featureless IVC-Bond phase map. The data, though, does show indications of vortices associated with IVC-A and IVC-B phases alone. These experimental results strongly suggest that strain plays a key role in selecting between different competitive correlated insulating ground states of MATBG.

We now compare the microscopic characteristics of the wavefunctions of the correlated insulating states at v = ±2 in different MATBG samples to theoretical predictions for these states. First, although the existence of an R3 pattern signifies IVC order, not all IVC candidate states display an R3 pattern. Particularly, the observation of an R3 pattern in all samples for these correlated insulators rules out the leading candidate Kramers intervalley coherent (K-IVC) state[13,31] at v = ±2, since despite being an IVC state, the symmetry properties of this state lead to the cancelation of any R3 spatial patterns in LDOS at zero magnetic field[10,11] (see SI). Recent theoretical studies[15,20,32] have proposed that in the presence of sufficient strain (ε > 0.1%), it is energetically favorable for the system to break moiré superlattice symmetries by forming an incommensurate Kekulé spiral (IKS) state. States with IKS order host IVC angles (i.e. the complex phase of our IVC order parameters) that exhibit a linear increase along a strain-dependent axis, incrementing by $\Delta\theta \neq 0$ between neighboring moiré unit cells. In Fig. 4a, we apply the same order parameter analysis we developed for understanding our experimental results to the theoretically calculated LDOS of an IKS state in a similar field of view as our experiments (computed for ε = 0.3%). IKS states appear moiré-periodic in IVC Total signal strength, while moiré translation and rotational symmetries appear broken in IVC-Bond, IVC-A and IVC-B maps, showing a stripe-like incommensurate state with $\Delta\theta \approx \frac{2\pi}{3}$. The pattern and direction of the stripe-like features depend in a complex manner on the degree and direction of the strain in the sample. Moreover, subtracting a linear background from these phase maps uncovers a lattice of vortices and antivortices with the same location and chirality across all three IVC order parameters, with their cores centered near zero-valued nodes of the total IVC strength (Fig. 4a). The remarkable resemblance of the moiré-periodic features of the magnitudes of the order parameters, the stripe-like features of their phases, and the lattice of vortex-antivortex features for the theoretically calculated IKS state (Fig. 4a) to those of our experimental results in strained samples (Fig. 3b) make IKS states the most likely candidates for the correlated insulating phases at $\nu = \pm 2$ in typical MATBG samples.

In contrast to samples of typical strain, ultra-low-strain samples show nearly moiré-periodic features in the LDOS of the correlated insulators at $\nu = -2$ (Fig. 3d), suggesting that IVC states that do not exhibit IKS order become more favorable in this limit. A state closely related to the K-IVC state, but often overlooked because it is not energetically favorable in existing numerical calculations[13], is the time-reversal symmetric intervalley coherent (T-IVC) state[10], whose wavefunction does host an R3 pattern at zero magnetic field. Fig. 4b shows an analysis of a calculation of LDOS for the T-IVC state, which shows moiré-periodic patterns in the IVC-Total, IVC-Bond, IVC-A, and IVC-B order parameters. Only the IVC-A and IVC-B phase maps show vortices, the presence, locations, and chiralities of which are expected due to the $C_3$-symmetry of the state (see SI). Overall, the experimental patterns of our extracted order parameters in our ultra-low-strain samples (Fig. 3d) are broadly consistent in their symmetries with a T-IVC state, therefore suggesting that a strained T-IVC state is perhaps a likely candidate phase for the correlated insulator phase of MATBG at $\nu = \pm 2$ in the low-strain limit.

Having identified the most likely candidate ground states for correlated insulating phases at $\nu = \pm 2$ in various samples of MATBG, we focus on how the wavefunction of the electronic states in MATBG evolves as we dope the system away from these integer fillings. In Figs. 5b-d, we show the evolution of the LDOS, as captured by our order parameter analysis, for a typical MATBG sample. An insulating gap at $\nu = +2$ closes with doping and reopens into a pseudogap phase between $\nu = +2$ and $+3$, upon which another correlated insulating phase appears in the same device at $\nu = +3$ (all measured at T = 200 mK). We also show order parameter maps for $\nu = -2.41$, at which the spectroscopic properties of our sample are consistent with the sample being superconducting[5] (same device region as in Fig. 3a, doped away from the $\nu = -2$ insulator). The wavefunction of the pseudogap phase at $\nu > +2$, the insulator at $\nu = +3$, and the superconductor at $\nu = -2.41$ all display R3 patterns in their wavefunctions, as captured by the strong moiré-periodic IVC Total signal over a wide field of view. However, we find that the IVC-Bond, IVC-A, and IVC-B order parameters show behavior distinct from that of the ν = ±2 correlated insulators (see SI for IVC-A/B). Doping away from ν = ±2 into either the superconducting or pseudogap phases results in a sudden change in the winding direction of IVC order parameter phases, the disappearance of the stripe-like patterns of the IVC order parameters, and the vanishing of associated vortices and antivortices observed across IVC order parameters at ν = ±2. These features persist to the ν = +3 insulating phase, whose order parameter maps show a strong resemblance to those of the adjacent pseudogap state. At this filling, our order parameter maps appear to show IVC phases that again increment by $\Delta\theta \neq 0$ between neighboring AA sites, possibly the result of an unpredicted form of IKS order, but such order is clearly distinguished from the IKS order appearing at ν = ±2 by the winding directions of IVC phases and the vortex configurations of these states.

Our findings shed new light on the relationship between the correlated insulators at ν = ±2 and the nearby superconductors in MATBG. Early transport studies have suggested that these phases are competing in nature[33–37], further supported by more recent STS measurements of the closing and subsequent reopening of the insulating gap at ν = -2 into a distinct, nodal tunneling gap in the superconducting state[5]. Here, our measurements demonstrate that although these states show locally very similar R3 patterns, subtle reorganizations of the LDOS on the graphene lattice-scale can result in dramatic changes in the characteristics of the IVC phases on the moiré-scale, which occur in a small region of doping between the insulator at ν = -2 and the superconductor at ν < -2. This suggests that the superconductor arises out of an IVC order that is qualitatively different from that of the neighboring insulator, which emphasizes important distinctions between the complex phase diagram of MATBG and those of other correlated superconductors, such as the high-$T_c$ cuprate superconductors[38].

Besides the correlated phases we have examined here, we expect our phase-sensitive techniques to be readily applied to identify the microscopic nature of topological phases in MATBG, such as those previously found in the presence of weak magnetic fields or for samples aligned with hBN[6–9,39,40]. Looking beyond graphene-based materials, intervalley coherent ground states are commonly predicted in other moiré material platforms[41,42], making our experimental approach relevant in the identification of such phases in these materials as well. In total, our work represents an unprecedentedly precise visualization of correlated many-body wavefunctions, demonstrating that even highly complex correlated quantum phases of matter can be established experimentally using the incisive capabilities of wavefunction mapping with STM.

## Methods

### Sample Preparation

The devices were fabricated using the 'tear-and-stack' method. Graphene and hBN are picked up with polyvinyl alcohol (PVA). Then, the heterostructure is flipped onto an intermediate structure consisting of methyl methacrylate co-polymer (Elvacite 2550 / transparent tape / Sylgard 184), and transferred to a prepatterned $SiO_2$ / Si chip with Ti / Au electrodes. Residual polymer is dissolved in N-methyl-2-pyrrolidone, dichloromethane, water, acetone, and isopropyl alcohol. We further clean the sample surface with AFM tip-cleaning and high-temperature forming gas annealing procedures. Finally, the device is annealed in ultrahigh vacuum at 170 °C for 12 hours and 400 °C for 2 hours before it is transferred into the STM.

### STM Measurement

STM / STS measurements were performed on a home-built dilution-refrigerator STM with tungsten tips prepared on a Cu(111) surface. The carrier density of MATBG was controlled by the gate voltage $V_g$ applied to degenerately doped Si, and the sample voltage $V_s$ applied to MATBG via the Au / Ti electrode. dI/dV was measured via lock-in detection of the A.C. tunneling current induced by an A.C modulation $V_{rms}$ added to $V_s$. Measurements are performed with sample bias voltages $V_s$ near zero to avoid inelastic tunneling due to K- or M-point phonons[43].

### Order Parameter Decomposition

For full details and explanations of this procedure, we refer readers to the SI. Briefly, large low-bias STM images are partitioned into smaller 0.25 - 1 $nm^2$ subregions. Each of these subregions is Fourier transformed with respect to the center of each subregion. We apply position-dependent phase factors to the FFT peaks to enforce a consistent origin across subregions. The three independent, complex values of each local FFT obtained at IVC wavevectors are decomposed into three complex IVC order parameters ("IVC Bond", "IVC Site A", and "IVC Site B"), which correspond to the three irreducible representations $\{(1, 1, 1), (1, \omega, \omega^2), \text{and } (1, \omega^2, \omega), \text{ where } \omega \equiv e^{2\pi i/3}\}$ of the $C_3$ point group. By construction, these order parameters are moiré-periodic if the LDOS is moiré-periodic.

### References:

1. Cao, Y. *et al.* Correlated insulator behaviour at half-filling in magic-angle graphene superlattices. *Nature* **556**, 80–84 (2018).

2. Cao, Y. *et al.* Unconventional superconductivity in magic-angle graphene superlattices. *Nature* **556**, 43–50 (2018).

3. Yankowitz, M. *et al.* Tuning superconductivity in twisted bilayer graphene. *Science* **363**, 1059–1064 (2019).

4. Lu, X. *et al.* Superconductors, orbital magnets and correlated states in magic-angle bilayer graphene. *Nature* **574**, 653–657 (2019).

5. Oh, M. *et al.* Evidence for unconventional superconductivity in twisted bilayer graphene. *Nature* **600**, 240–245 (2021).
6. Sharpe, A. L. *et al.* Emergent ferromagnetism near three-quarters filling in twisted bilayer graphene. *Science* **365**, 605–608 (2019).
7. Serlin, M. *et al.* Intrinsic quantized anomalous Hall effect in a moiré heterostructure. *Science* **367**, 900–903 (2020).
8. Nuckolls, K. P. *et al.* Strongly correlated Chern insulators in magic-angle twisted bilayer graphene. *Nature* **588**, 610–615 (2020).
9. Xie, Y. *et al.* Fractional Chern insulators in magic-angle twisted bilayer graphene. *Nature* **600**, 439–443 (2021).
10. Călugăru, D. *et al.* Spectroscopy of Twisted Bilayer Graphene Correlated Insulators. *Phys. Rev. Lett.* **129**, 117602 (2022).
11. Hong, J. P., Soejima, T. & Zaletel, M. P. Detecting Symmetry Breaking in Magic Angle Graphene Using Scanning Tunneling Microscopy. *Phys. Rev. Lett.* **129**, 147001 (2022).
12. Kang, J. & Vafek, O. Strong Coupling Phases of Partially Filled Twisted Bilayer Graphene Narrow Bands. *Phys. Rev. Lett.* **122**, 246401 (2019).
13. Bultinck, N. *et al.* Ground State and Hidden Symmetry of Magic-Angle Graphene at Even Integer Filling. *Phys. Rev. X* **10**, 031034 (2020).
14. Lian, B. *et al.* Twisted bilayer graphene. IV. Exact insulator ground states and phase diagram. *Phys. Rev. B* **103**, 205414 (2021).
15. Kwan, Y. H. *et al.* Kekulé Spiral Order at All Nonzero Integer Fillings in Twisted Bilayer Graphene. *Phys. Rev. X* **11**, 041063 (2021).
16. Tarnopolsky, G., Kruchkov, A. J. & Vishwanath, A. Origin of Magic Angles in Twisted Bilayer Graphene. *Phys. Rev. Lett.* **122**, 106405 (2019).
17. Song, Z.-D. & Bernevig, B. A. Magic-Angle Twisted Bilayer Graphene as a Topological Heavy Fermion Problem. *Phys. Rev. Lett.* **129**, 047601 (2022).

18. Liu, X. *et al.* Visualizing broken symmetry and topological defects in a quantum Hall ferromagnet. *Science* **375**, 321–326 (2022).
19. Coissard, A. *et al.* Imaging tunable quantum Hall broken-symmetry orders in graphene. *Nature* **605**, 51–56 (2022).
20. Wagner, G., Kwan, Y. H., Bultinck, N., Simon, S. H. & Parameswaran, S. A. Global Phase Diagram of the Normal State of Twisted Bilayer Graphene. *Phys. Rev. Lett.* **128**, 156401 (2022).
21. Lake, E., Patri, A. S. & Senthil, T. Pairing symmetry of twisted bilayer graphene: A phenomenological synthesis. *Phys. Rev. B* **106**, 104506 (2022).
22. Khalaf, E., Chatterjee, S., Bultinck, N., Zaletel, M. P. & Vishwanath, A. Charged skyrmions and topological origin of superconductivity in magic-angle graphene. *Sci. Adv.* **7**, eabf5299 (2021).
23. Kozii, V., Isobe, H., Venderbos, J. W. F. & Fu, L. Nematic superconductivity stabilized by density wave fluctuations: Possible application to twisted bilayer graphene. *Phys. Rev. B* **99**, 144507 (2019).
24. Christos, M., Sachdev, S. & Scheurer, M. S. Superconductivity, correlated insulators, and Wess–Zumino–Witten terms in twisted bilayer graphene. *Proc. Natl. Acad. Sci.* **117**, 29543–29554 (2020).
25. Wong, D. *et al.* A modular ultra-high vacuum millikelvin scanning tunneling microscope. *Rev. Sci. Instrum.* **91**, 023703 (2020).
26. Kerelsky, A. *et al.* Maximized electron interactions at the magic angle in twisted bilayer graphene. *Nature* **572**, 95–100 (2019).
27. Choi, Y. *et al.* Electronic correlations in twisted bilayer graphene near the magic angle. *Nat. Phys.* **15**, 1174–1180 (2019).
28. Jiang, Y. *et al.* Charge order and broken rotational symmetry in magic-angle twisted bilayer graphene. *Nature* **573**, 91–95 (2019).

29. Rutter, G. M. *et al.* Scattering and Interference in Epitaxial Graphene. *Science* **317**, 219–222 (2007).

30. Wong, D. *et al.* Cascade of electronic transitions in magic-angle twisted bilayer graphene. *Nature* **582**, 198–202 (2020).

31. Yu, J. *et al.* Spin skyrmion gaps as signatures of intervalley-coherent insulators in magic-angle twisted bilayer graphene. Preprint at https://doi.org/10.48550/arXiv.2206.11304 (2022).

32. Wang, T. *et al.* Kekul\'e spiral order in magic-angle graphene: a density matrix renormalization group study. Preprint at https://doi.org/10.48550/arXiv.2211.02693 (2022).

33. Stepanov, P. *et al.* Untying the insulating and superconducting orders in magic-angle graphene. *Nature* **583**, 375–378 (2020).

34. Saito, Y., Ge, J., Watanabe, K., Taniguchi, T. & Young, A. F. Independent superconductors and correlated insulators in twisted bilayer graphene. *Nat. Phys.* **16**, 926–930 (2020).

35. Arora, H. S. *et al.* Superconductivity in metallic twisted bilayer graphene stabilized by WSe2. *Nature* **583**, 379–384 (2020).

36. Cao, Y. *et al.* Nematicity and competing orders in superconducting magic-angle graphene. *Science* **372**, 264–271 (2021).

37. Liu, X. *et al.* Tuning electron correlation in magic-angle twisted bilayer graphene using Coulomb screening. *Science* **371**, 1261–1265 (2021).

38. Comin, R. & Damascelli, A. Resonant X-Ray Scattering Studies of Charge Order in Cuprates. *Annu. Rev. Condens. Matter Phys.* **7**, 369–405 (2016).

39. Wu, S., Zhang, Z., Watanabe, K., Taniguchi, T. & Andrei, E. Y. Chern insulators, van Hove singularities and topological flat bands in magic-angle twisted bilayer graphene. *Nat. Mater.* **20**, 488–494 (2021).

40. Saito, Y. *et al.* Hofstadter subband ferromagnetism and symmetry-broken Chern insulators in twisted bilayer graphene. *Nat. Phys.* **17**, 478–481 (2021).

41. Devakul, T., Crépel, V., Zhang, Y. & Fu, L. Magic in twisted transition metal dichalcogenide bilayers. *Nat. Commun.* **12**, 6730 (2021).

42. Xie, Y.-M., Zhang, C.-P. & Law, K. T. Topological p_x + ip_y inter-valley coherent state in Moiré MoTe2 / WSe2 heterobilayers. Preprint at https://doi.org/10.48550/arXiv.2206.11666 (2022).

43. Natterer, F. D. *et al.* Strong Asymmetric Charge Carrier Dependence in Inelastic Electron Tunneling Spectroscopy of Graphene Phonons. *Phys. Rev. Lett.* **114**, 245502 (2015).

**Acknowledgements**

We thank Oskar Vafek, Xiaomeng Liu, Cheng-Li Chiu, and Gelareh Farahi for helpful discussions. We thank Hao Ding for helpful technical discussion. This work was primarily supported by the Gordon and Betty Moore Foundation's EPiQS initiative grants GBMF9469 and DOE-BES grant DE-FG02-07ER46419 to A.Y. Other support for the experimental work was provided by NSF-MRSEC through the Princeton Center for Complex Materials NSF-DMR-2011750, NSF-DMR-1904442, ARO MURI (W911NF-21-2-0147), and ONR N00012-21-1-2592. T.S. was supported by a fellowship from Masason foundation, and by the U.S. Department of Energy, Office of Science, National Quantum Information Science Research Centers, Quantum Systems Accelerator. J.P.H. was supported by the Princeton University Department of Physics. M.P.Z. was supported by the U.S. Department of Energy, Office of Science, Office of Basic Energy Sciences, Materials Sciences and Engineering Division, under Contract No. DE-AC02-05CH11231, within the van der Waals Heterostructures Program (KCWF16), and the Alfred P Sloan Foundation. D.C., B.A.B. and N.R. were supported by the European Research Council (ERC) under the European Union's Horizon 2020 research and innovation programme (grant agreement No. 101020833), the ONR Grant No. N00014-20-1-2303, Simons Investigator Grant No. 404513, the Gordon and Betty Moore Foundation through the EPiQS Initiative, Grant GBMF11070 and Grant No. GBMF8685, NSF-MRSEC Grant No. DMR-2011750, BSF Israel US foundation Grant No. 2018226, and the Princeton Global Network Funds. J.H.A. was supported by a Hertz Fellowship. N.R. acknowledges support from the QuantERA II Programme that has received funding from the European Union's Horizon 2020 research and innovation programme under Grant Agreement No 101017733. K.W. and T.T. acknowledge support from the Elemental Strategy Initiative conducted by the MEXT, Japan, grant JPMXP0112101001, JSPS KAKENHI grant 19H05790 and JP20H00354.

**Author Contributions**

K.P.N., R.L.L., M.O., D.W. and A.Y. designed the experiment. D.W., K.P.N., M.O., and R.L.L. fabricated the devices used for the study. M.O., R.L.L., K.P.N., and D.W. carried out STM / STS measurements. T.S., J.P.H., and M.P.Z. designed the order parameter decomposition scheme and built the software suite capable of performing this decomposition on STM data, with input from all authors. R.L.L., D.W., K.P.N., and M.O. performed the data analysis using this software suite and maintained the code. T.S., J.P.H., D.C., and J.H.A. performed simulations of the LDOS of candidate insulating ground states under the guidance of B.A.B., N.R., and M.P.Z. K.W. and T.T. synthesized the hBN crystals. All authors discussed the results and contributed to the writing of the manuscript.

**Data Availability**

The data that supports the findings of this study are available from the corresponding author upon reasonable request.

**Competing Interests**

The authors declare no competing interests.

# Figure 1

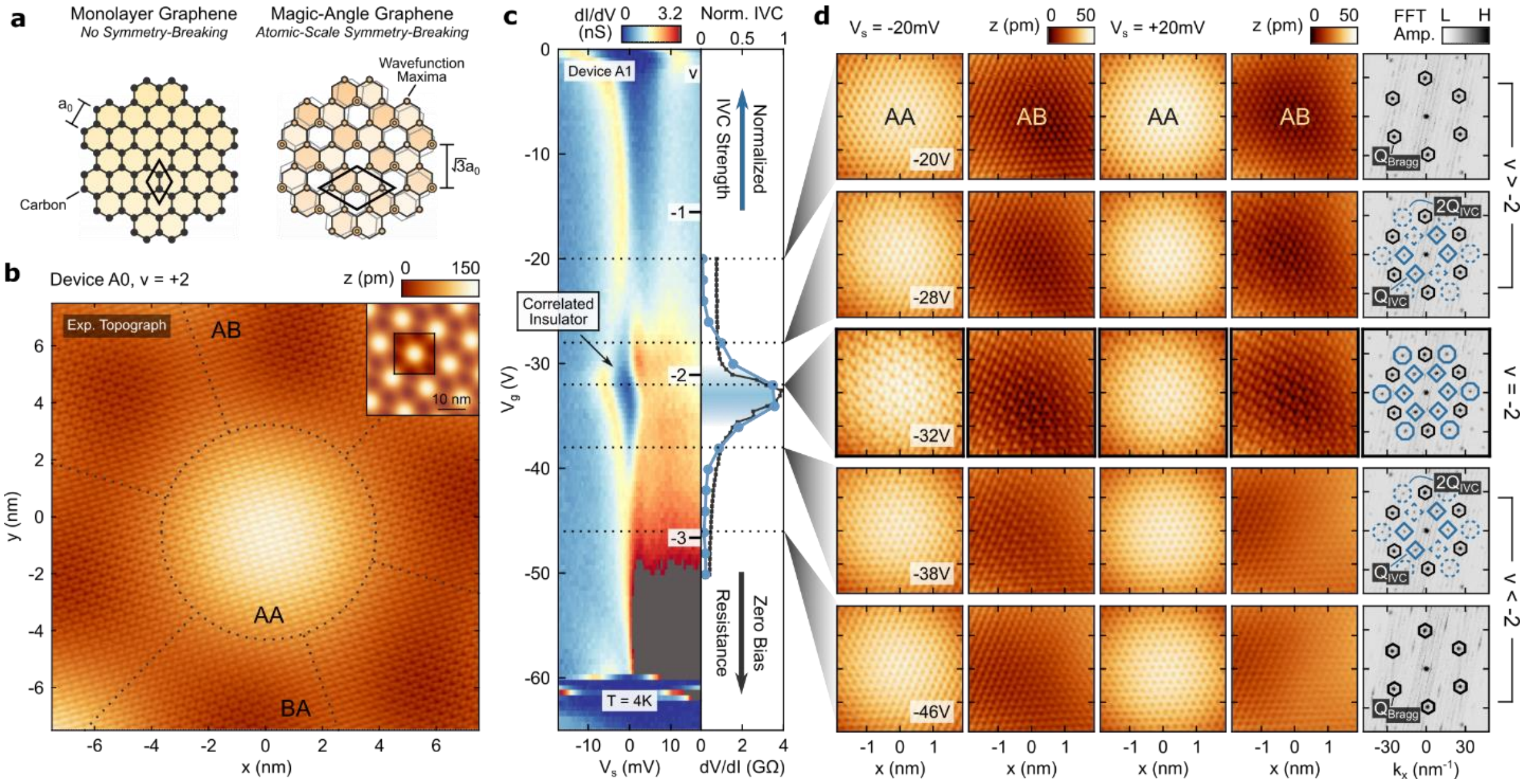

**Figure 1 | Imaging Atomic-Scale Kekulé Patterns in MATBG. a,** Schematic diagram of the underlying graphene atomic lattice and the (√3 x √3)R30° “R3 pattern” in magic-angle graphene, which triples the graphene unit cell. **b,** STM topographic image of a magic-angle region (Device A0) at T = 4 K, B = 0 T, and v = +2. The dashed circle surrounds the AA region, while radial dotted lines indicate the bridge regions that separate AB from BA regions. Inset: Zoomed-out image of the device. The black box shows where **b** was obtained. **c,** dI/dV($V_s$, $V_g$) (left) measured at the center of an AA site at T = 4 K and B = 0 T. The deep suppression near v = -2 is indicative of an incipient correlated insulating phase. Shown on the right is the normalized IVC strength $\Sigma|FFT(Q_{IVC})| \,/\, \Sigma|FFT(Q_{Bragg})|$ and the zero-bias resistance dV/dI($V_s$ = 0 V) as a function of $V_g$. **d,** Topographic maps ($V_s$ = ±20 mV) and Fourier transforms (FFT) of topographic maps ($V_s$ = -20 mV) at $V_g$ = -20 V (top), -28 V (upper middle), -32 V (middle), -38 V (lower middle), and -46 V (bottom). Maps at $V_g$ = -20 V and -46 V were obtained outside of the v = -2 insulating gap, and show no R3 pattern. Maps at $V_g$ = -32 V were obtained in the v = -2 insulating gap, and show a strong R3 pattern across the moiré superlattice. FFTs are derived from 40 x 40 nm$^2$ images obtained at $V_s$ = -20 mV, from which the subregions shown in **d** were taken (see SI for full images). FFTs show the reciprocal lattice vectors ($Q_{Bragg}$; black hexagon markers) of the graphene atomic lattice. As the gate voltage is decreased, a new set of peaks emerges that corresponds to the periodicity of the R3 tripled unit cell ($Q_{IVC}$ and $2Q_{IVC}$; blue diamond and octagon markers). Dashed diamonds and octagons correspond to peaks that are faint but still present. See Supplementary Information for tunneling parameters and full datasets.

# Figure 2

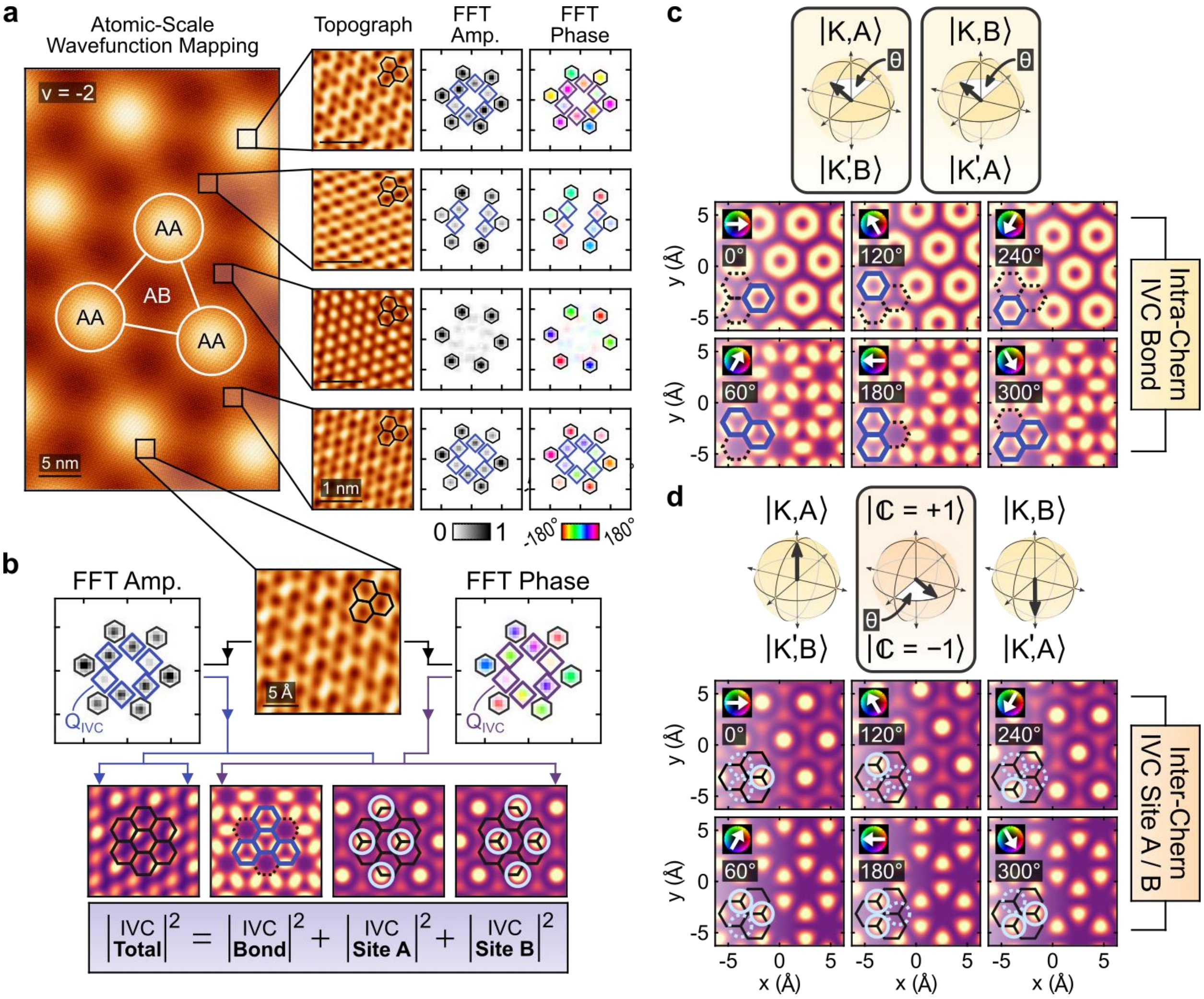

**Figure 2 | Symmetry-Based Order Parameter Decomposition and Inter-Chern-Sector Coherence. a,** Cropped large-scale (50 x 30 nm$^2$) atomically resolved STM topographic image obtained at v = -2. Small-scale (2 x 2 nm$^2$) zoomed-in images have been FFT filtered around the $Q_{Bragg}$ and $Q_{IVC}$ wavevectors, as shown in local FFTs of these regions on the right. **b,** Schematic diagram of the local FFT order parameter decomposition. Small-scale (0.25 - 1 nm$^2$) topographic images, cropped from the large-scale image in **a**, are used to extract local FFT amplitude (left) and phase (right) information. Information from the FFT at the $Q_{IVC}$ wavevectors is further decomposed into four IVC order parameters (bottom row). **c,d,** Representative simulations of STM images that illustrate purely IVC Bond (**c**) and IVC Site A / B **(d)** order parameters. Order parameter phase increments of 60° are plotted to illustrate the order parameters' connection to the new R3 sublattice degree of freedom. Images that differ in IVC angle by 120° only differ by a translational shift of one graphene UC. Images that differ in IVC angle by 180° add in LDOS to produce patterns that do not show a tripled graphene UC. Yellow Bloch spheres (top of **c**) depict the chiral model Chern decomposition, which identifies a four-fold degeneracy of Chern subbands within each Chern sector. The orange Bloch sphere (top of **d**) captures the inter-Chern-sector coherence between two Chern sectors. This Chern Bloch sphere interferes wavefunctions from different Chern sectors.

# Figure 3

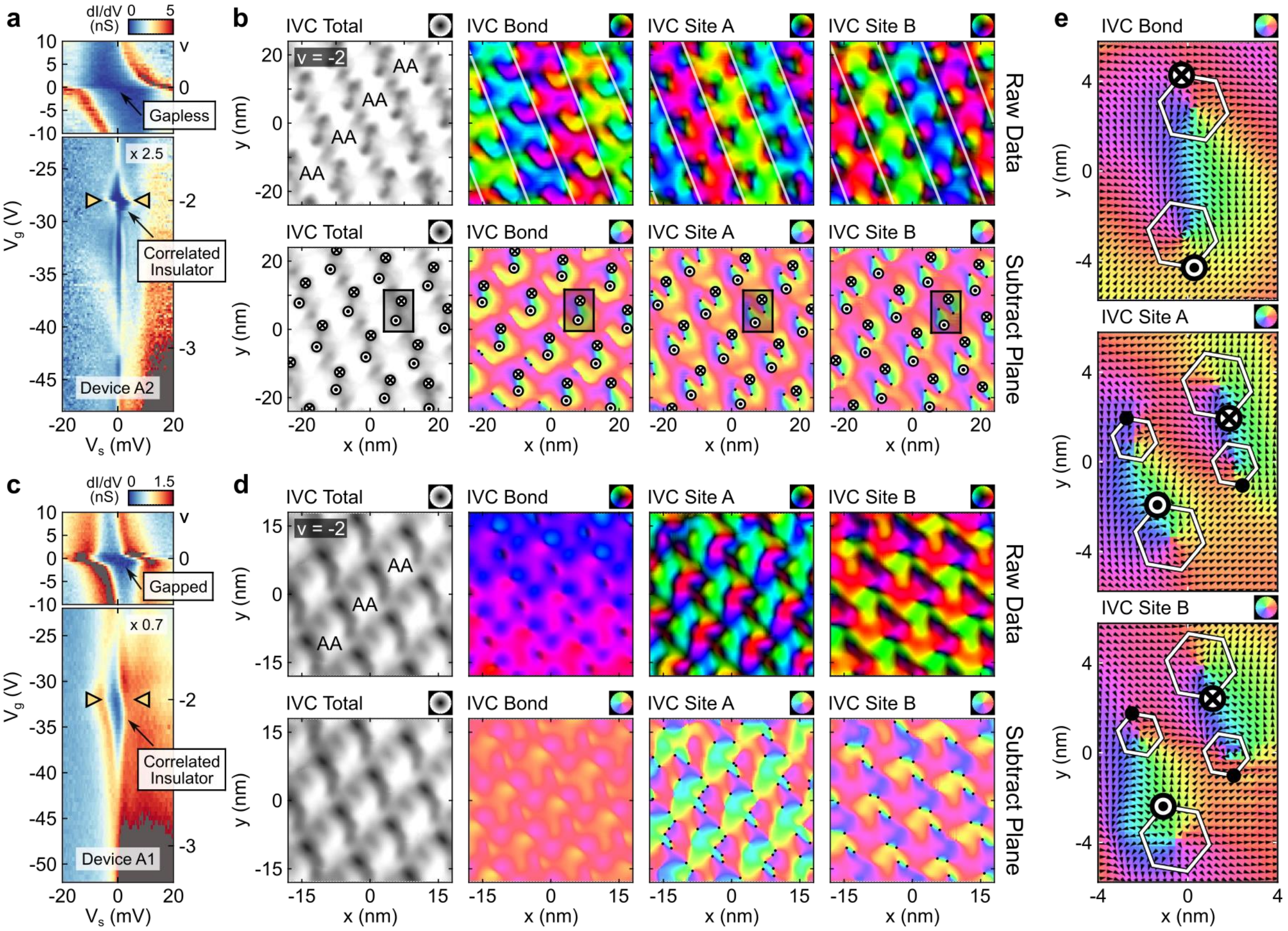

**Figure 3 | Distinguishing Correlated Insulators at v = ±2 via Moiré Translation Symmetry and IVC Isospin Vortices. a,** dI/dV($V_s$, $V_g$) measured at the center of an AA site at T = 200 mK and B = 0 T near v = 0 (top) and near v = -2 and -3 (bottom) in a gapless CNP device region. Insulating spectra occur at v = -2, and gapless semimetallic spectra occur near v = 0. Superconducting and pseudogap spectra occur between v = -2 and -3. Yellow arrows indicate the density at which images in **b** were obtained. **b,** Local order parameter decomposition of a roughly 50 x 50 $nm^2$ low-bias STM image of the v = -2 correlated insulating state, obtained at $V_s$ = -5 mV. "IVC Total" measures the intensity of the R3 pattern, while "IVC Bond / Site A / Site B" measure the phase of the R3 pattern. The top row shows raw order parameter plots, and the bottom row shows vortex-annotated order parameter plots without amplitude information, where a linear phase gradient has been subtracted from the three IVC phase plots on the right to highlight the IVC vortices (into and out-of-plane markers and black dot markers). Black shaded boxes represent the locations where plots in **e** were obtained. **c,** dI/dV($V_s$, $V_g$) measured at the center of an AA site at T = 4 K and B = 0 T near v = 0 (top) and near v = -2 and -3 (bottom) in a gapped CNP device region. Insulating spectra occur both at v = -2 and v = 0. **d,** Same as **b**, for v = -2 in this gapped CNP device region in a roughly 40 x 40 $nm^2$ region, obtained at $V_s$ = -20 mV. **e,** Zoom-in plots of **b**, illustrating the chirality of the observed IVC vortices. Each small black arrow represents the phase of the IVC order parameter at that point.

# Figure 4

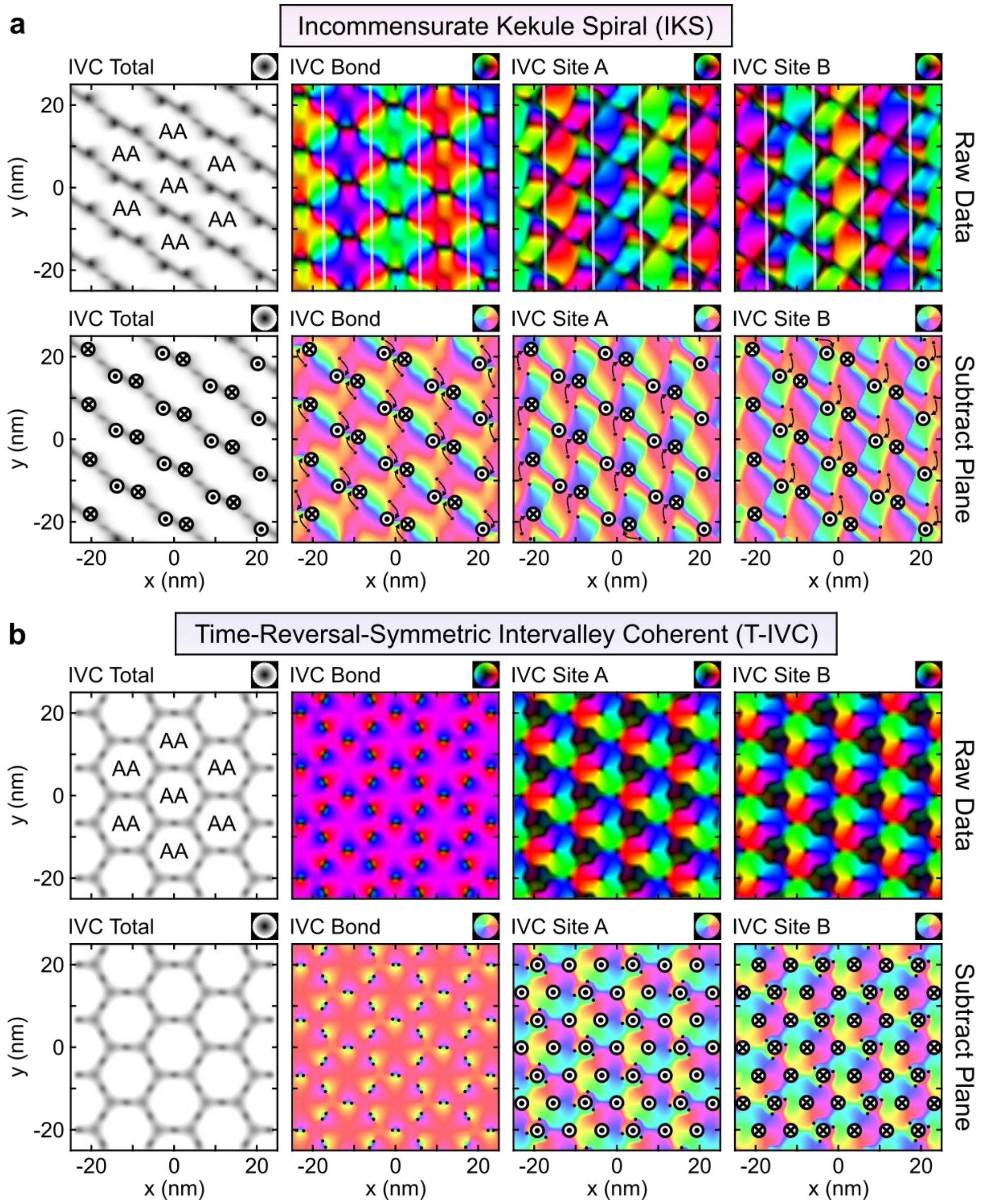

**Figure 4 | Candidate Theoretical Ground States. a,** Local order parameter decomposition performed on LDOS simulations of an Incommensurate Kekulé Spiral (IKS) candidate state. IKS states break moiré translation symmetry, where neighboring stripes of AA sites differ by $\Delta\theta \approx \frac{2\pi}{3}$ in IVC Bond / Site A / Site B order parameters. IKS states support a moiré-scale lattice of IVC vortex-antivortex pairs, one per moiré unit cell, correlated across all IVC order parameters. Black arrows point to vortex locations covered by neighboring vortex markers. **b,** Same as **a** for the Time-Reversal-Symmetric Intervalley Coherent (T-IVC) candidate state. The T-IVC state preserves moiré translation symmetry, where all AA sites show identical features in all IVC order parameters.

# Figure 5

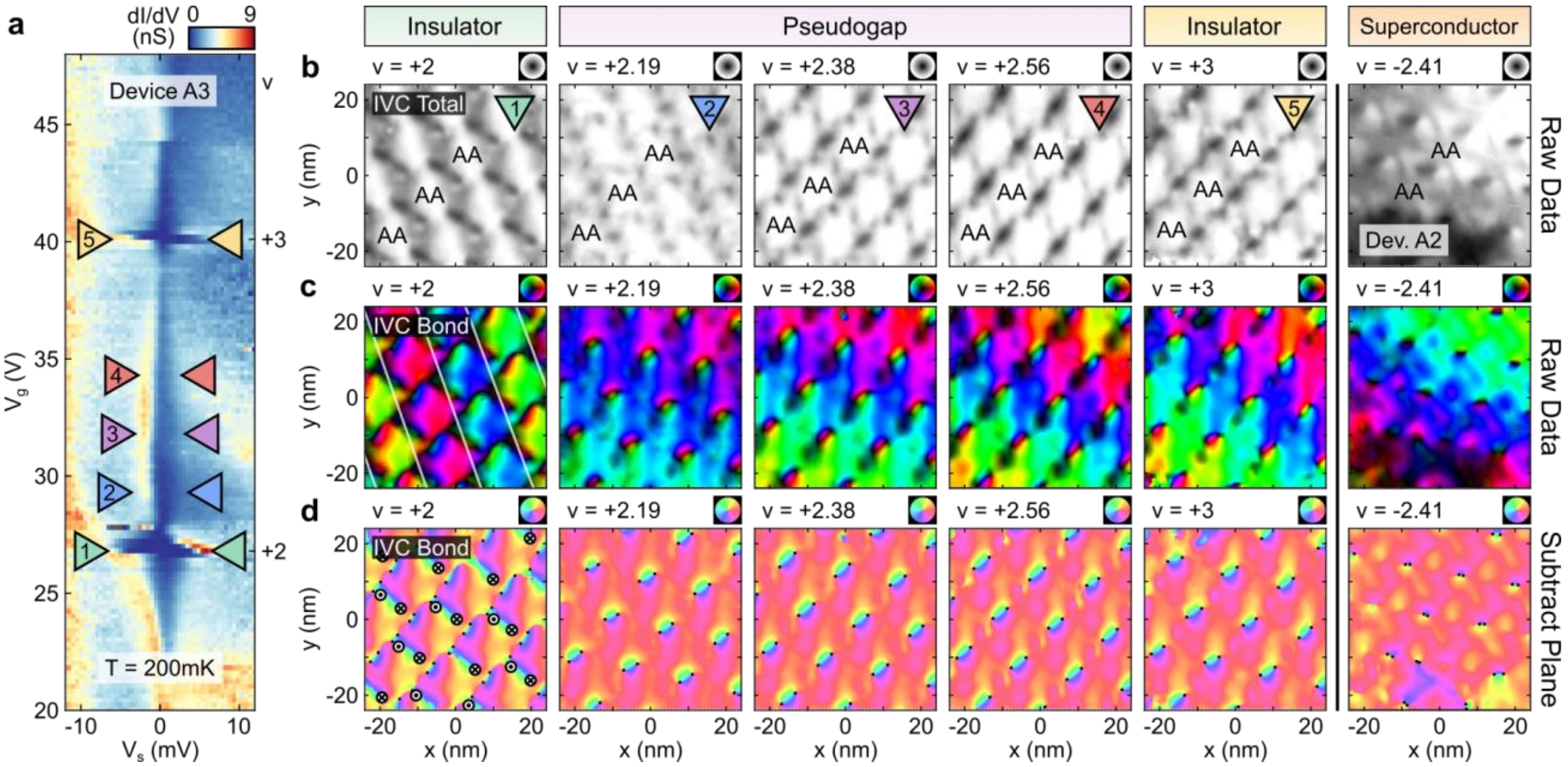

**Figure 5 | Doping Correlated Insulators into Superconducting and Pseudogap Phases. a,** dI/dV($V_s$, $V_g$) measured at the center of an AA site at T = 200 mK and B = 0 T between v = +2 and +3 in a gapless CNP device region. Insulating spectra occur at v = +2 and v = +3, while pseudogap spectra occur between these fillings. Colored arrow markers correspond to images in **b**, **c**, and **d**. **b,** Raw IVC Total order parameters extracted from STM images obtained in insulating phases (v = +2, +3), pseudogap phases (v = +2.19, +2.38, +2.56), and superconducting phases (v = -2.41 from Device A2, see Fig. 3a for STS measurements). **c,** Same as **b**, for the Raw IVC Bond order parameters. **d,** Same as **b**, for the IVC Bond order parameters with a linear phase gradient removed to emphasize the observed IVC vortex configurations.